\documentstyle[12pt,epsfig]{article}
\author{Pedro BLECUA, 
\footnote{402535@docto.unizar.es}
\ Luis J. BOYA 
\footnote{luisjo@unizar.es}
\ and Antonio SEGUI
\footnote{segui@unizar.es}\\
Departamento de F\'{\i}sica Te\'orica, Facultad de Ciencias\\
Universidad de Zaragoza.- E-50009 ZARAGOZA, Spain}
\title{New solvable quantum mechanical potentials by iteration of the free $V=0$ potential}
\begin{document}
\maketitle
\begin{abstract}

A huge family of solvable potentials can be generated by systematically exploiting the factorization (Darboux) method. Staarting from the free case, a large class of the known solvable families is thus reproduced, together with new ones. We explicitly find and solve several new singular potentials obtained by iteration from the $V=0$ case; some of them have an $E=0$ bound state and constant phase shift without being explicitly scale invariant. The new potentials are rational functions, and can be related to rational solutions of the KdV family.
 
\end{abstract}
\vspace{1.5cm}
PACS \# 03.65 (Fd,Ge,Nk)
\newpage
\newcommand{\be}{\begin{equation}}
\newcommand{\ee}{\end{equation}}
\newcommand{\la}{\label}
\renewcommand{\thesection}{\arabic{section}}
\renewcommand{\theequation}{\thesection.\arabic{equation}}

\section{Introduction}

The search for solvable potentials in quantum mechanics is an old and large industry. In this paper we use the factorization method of Darboux (\cite{1},\cite{2})
 to formally construct infinite families of fully solvable potentials, all related to the free $V(x)=0$ case.

The essence of the method is the following. We start from a hamiltonian $H_0$ and a particular solution ($E_0, \phi_0$) ($D \equiv {d \over dx}$ will be used troughout the paper) 
\be\la{1} 
H_0=-D^2+V_0(x); \qquad H_0\phi_0(x)=E_0\phi_0(x)
\ee

The solution $\phi_0$ \it needs  not to be physical, i.e. 
\rm it might blow up at finite or infinite distances. Then we construct the partner potential

\be\la{2}
V_1(x)-E_0=W'^2(x)+W''(x), \qquad with \  \phi_o(x)\equiv \exp(-W(x))
\ee
such that the new hamiltonian $H_1=-D^2+V_1(x)$ has as solutions 

\be\la{3}
H_1\psi_k=E_k\psi_k \quad where \  \psi_k=A\phi_k, \ A=D+W'(x) \ and \  H_0\phi_k=E_k\phi_k
\ee

In other words, all the solutions ($\phi_k, E_k$) of the first problem generate solutions ($A\phi_k, E_k$) of the new problem. But as $A\phi_0=0$, the new solution is $\psi_0=\phi_o^{-1}$. At times, some solutions have to be excluded by physical reasons.

We shall need also the second solution from the known one $\psi_1$ at the same energy; it is 

\be\la{4}
\psi_2=\psi_1\int \psi_1^{-2} dx
\ee

These resuls are fairly well-known; for convenience of the reader we supply simple proofs of the above statements in Appendix {I}.

Our program is to start with the zero potential (free particle) $V(x)=V_0=0$ and iterate new potentials from its solutions. There are \it four 
 \rm types of different potentials, from four solutions as follows: \cite{3}

\begin{eqnarray}\la{5}
(E=k^2>0, k=1): \qquad  &\phi&=\cos(x), \Rightarrow V_1(x)=+2\sec^2(x) \nonumber\\
(E=0): \qquad  &\phi&=x, \Rightarrow V_1(x)={2 \over x^2}\nonumber\\
(E=-k^2<0,k=1): \qquad  &\phi&=\cosh(x), \Rightarrow V_1(x)=-2\cosh^{-2}(x)\nonumber\\
&\phi&=\sinh(x), \Rightarrow V_1(x)=+2\sinh^{-2}(x)
\end{eqnarray}

Notice the last three ``solutions'' are unphysical; a fifth potential $V_1(x)=2\csc^2(x)$ is just the first one displaced ${\pi \over 2}$.

The complete solutions for these \it first-step
\rm potentials are obtained by pulling from the free solution by the corresponding $A$ operator, see (\ref{3}). Here we recall only the situation for one of the more important cases, providing the only case with a potential valid in the whole straight line $(-\infty,+\infty)$ :
\be\la{6}
V(x)=-2\cosh^{-2}(x), \qquad W'(x)=-\tanh(x)
\ee
There is a \it ground state

\be\la{7}
\psi_0=\phi_0^{-1}(x)=\cosh^{-1}(x) \qquad (unnormalized) 
\ee
\rm and scattering solutions
\be\la{8}
\psi_k(x)=(D-\tanh(x))\exp(ikx)=(ik-\tanh(x))\exp(ikx)
\ee
corresponding to a transparent (reflectionless) potential, with pure transmission
\be\la{9}
t(k)= \ transmission \ ={ik-1 \over ik+1}
\ee

This potential is \it critical,
\rm having an $E=0$ resonance; in fact, all transparent potentials are critical \cite{4}
; our potential (\ref{6}) corresponds to a single soliton.     
\section{Study of a second-step potential}

The power of the method can be seen now; as the $V(x)=0$ case is trivially solvable for any energy $E$, physical or unphysical, we have now four potentials, all fully solvable; and from \it each
\rm solution of \it each 
\rm energy of \it each
\rm potential, we can in principle obtain a new, still fully solvable potential. In this paper we shall elaborate only in the families associated to the $E=0$ (intermediate) case.

We start now from the ``centrifugal'' potential $V(x)={2 \over x^2}$ and recall the (unphysical!) solutions: one is ${1 \over x}$ , as $\phi_0(x)=x$ is the starting solution for the $V(x)=0$, $E=0$ case; the other is (cf. (\ref{4})) ${1 \over x}\int x^2 dx \sim x^2$. Although both solutions blow up at $x=0$ and $x=\infty$ respectively, they are instrumental in obtaining a one-parameter family of \it bona-fide
\rm, physical potentials:

From the general $E=0$ wavefunction $\phi(x)={a \over x}+bx^2$, with $W'(x)=-{\phi'(x) \over \phi(x)}$ and $\mu\equiv {a \over b}>0$ we get the new, second-step potential family
\be\la{10}
V_\mu(x)=W'^2(x)+W''(x)={6x(x^3-2\mu) \over (x^3+\mu)^2}
\ee

The new potential(s) is singular: it has a \it double pole
\rm at $x=-c \ , \ c\equiv \mu^{1/3} \ , \ 0<\mu<\infty$. For $x>-c$ it has an atractive part, and a repulsive tail; for $x<-c$ is purely repulsive; both iterpolate between
\be\la{11}
V(x)\sim {2 \over (x+c)^2} \ (x\simeq -c) \ ... \ V(x)\sim {6 \over \left | x\right | ^2} \ ({x \to \pm \infty})
\ee
See Fig.1
\begin{figure}
\begin{center}
\includegraphics[width=13cm]{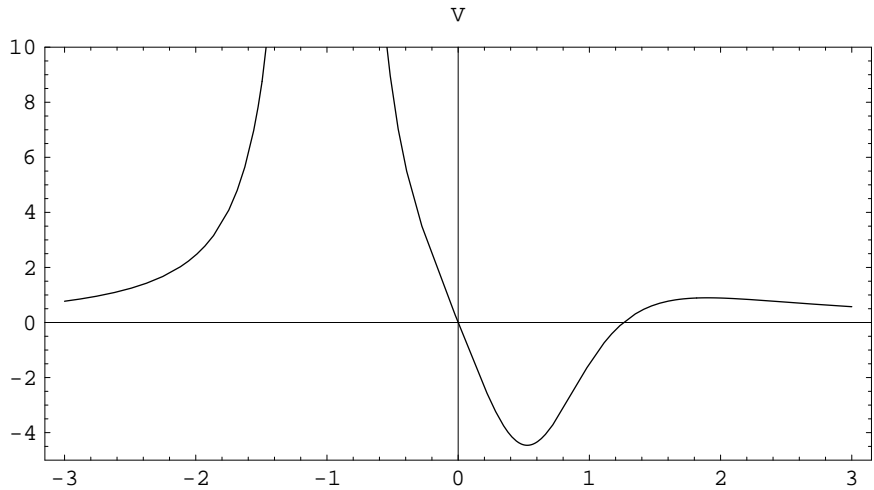}
\end{center}
\caption{{The potential (2.10) , with two independent parts ($\mu \equiv 1$)}}
\label{Figure 1}
\end{figure}

Both $V_1={2 \over x^2}$ and our potential (\ref{10}) correspond to some rational solutions of the KdV equation \cite{5};
  the relation is interesting and we elaborate on it in Appendix {II}.

Of course, the barrier at $x=-c$ is impenetrable: we have two different physical problems.

\bf 1) Case $x\geq -c$. 
\rm A \it bound state
\rm  candidate with $E=-\kappa^2<0$ would behave like $\exp (-\kappa x)$ at large $x$, so we try the $E<0$ (unphysical) solution $\phi(x)=\exp(-\kappa x)$ of the $V(x)=0$ case and \it prolongate it twice
\rm , as explained above. We find
\begin{eqnarray}\la{12}
\psi_0(x)=A_2A_1\phi(x) &=& (D-{2x^3-\mu \over x(x^3+\mu)})(D-{1 \over x})\exp(-\kappa x) \nonumber\\
&=& (\kappa^2+3x{1+\kappa x \over \mu+x^3})\exp(-\kappa x)
\end{eqnarray}

For this wavefunction to be physical it has to be zero at the singularity: this leads to the eigenvlue equation $1+\kappa (x=-c)=0$. Hence there is a \it single
\rm bound state with energy $E=-\kappa^2 \ , \ \kappa ={1/c}=\mu^{-1/3}$ , and whose (un-)normalized wavefunction is
\be\la{13}
\psi_0(x)=({1/c^2}+{{3x/c} \over x^2-cx +c^2})\exp(-{x/c})={({x/c}+1)^2 \over x^2-cx+c^2}\exp(-{x/c})
\ee
which behaves in the expected way for a ground state: nodeless, normalizable, decaying fast at ${x \to \infty}$. By construction, this state is \it the only
\rm bound state.

For $E=k^2>0$ we have total reflection; we write the wavefunction as
\be\la{14}
\psi_k(x)=A_2A_1(a\exp(ikx)+b\exp(-ikx))
\ee
and impose $\psi_k(x=-c)=0$; this fixes ${a/b}$ as a phase,
\be\la{15}
{a/b}=-{1-ikc \over 1+ikc}\exp(2ikc)
\ee

From the asymptotic behaviour we extract the S-matrix as usual in scattering in  one radial dimension
\be\la{16}
\psi_k(x>>0)\equiv \exp(-ik(x+c))-S(k)\exp(ik(x+c))
\ee
and comparing with (\ref{14}), we derive
\be\la{17}
S(k)={1-ikc \over 1+ikc}
\ee
Or, for the phase shift $S(k)\equiv \exp(2i\delta(k))$
\be\la{18}
\delta(k)=-\arctan(kc)\quad mod \ \pi
\ee
wich has to be interpreted carefully: with centrifugal tails ${V(x) \to {\lambda \over x^2}} \ , for \ x>>$ the usual rule $\delta(\infty)=0$ does not apply. The interpretation of (\ref{18}) is as follows:

At $k=0$ , the bound state contributes $+\pi$ to the phase shift (Levinson's theorem) and the \it long tail
\rm ($x>>0$) of the potential, which is ${6/x^2}\equiv {l(l+1)/x^2} \ (l=2)$ , contributes $-2({\pi/2})$; hence, $\delta(k=0)=0$ . At very large $k$, the phase shift is dominated only by the short tail, still centrifugal $+{2/(x+c)^2}$, which should produce a $-{\pi/2}$ shift. All this is reproduced by (\ref{18}) with the determination $\arctan(0)=0$ \footnote{ It is well known, \it e.g.
\rm in $3$D scattering, that a purely centrifugal potential $V_{cent}={l(l+1)/x^2}$ produces a negative constant phase shift $\delta_{cent}(k)=-l{\pi /2}$}.     

Notice the S-matrix (\ref{17}) is about the simplest with the pole at the bound state $k=+{i/c}$ : this is very similar to the forward amplitude for the solitonic scattering (\ref{9}): it seems that the fact that there is a \it single
\rm bound state determines the phase shift, and other features of the potential are somehow irrelevant.

\bf 2) Case $x<-c$. 
\rm Here there is also total reflection, but obviously no bound state, and an analogous calculation gives  the S-matrix as \it inverse
\rm of the previous one, and we get
\be\la{19}
\delta(k)=+\arctan(kc) \quad mod \ \pi
\ee

At $k=0$ the long tail contributes $-2({\pi/2})$, hence we determine $\arctan(0)=-\pi$ ; as ${k \to \infty}$ , the short tail dominates with $\delta(\infty)=-{\pi/2}$ ; of course, the only invariant statement is the difference, that is, the span $\Delta \equiv \delta(0)-\delta(\infty)$.

A surprising property of the potential (\ref{10}) has to do with the \it golden ratio
\rm $\Phi \equiv {(1+\sqrt{5})/2}$ : for $x>0$ , the maximum $x_M$ and minimum $x_m$ of $V(x)$ in (\ref{10}) are
\be\la{20}
x_m^3=(2+3\Phi), \qquad x_M^3=(2-{3/\Phi})=x_m^3 \ ({\Phi \to -{1/\Phi}})
\ee
and the same happens for the values of the potential:
\be\la{21}
V(x_m)=2\Phi{(2+3\Phi)^{1/3} \over (1+\Phi)^2}, \qquad V(x_M)=V(x_m) \ ({\phi \to -{1/\Phi}})
\ee

While we do not fully understand this relation, we notice the same thing appears in the KdV for two solitons with velocities $k_1$ and $k_2$: $\Phi={k_2/k_1}$, separates the overlapping and non-overlapping profiles (\cite{5}, p. 190; the discovery seems due to Lax)
; it is another intriguing connection between solitons and special potentials.

\section{Some generalizations}
\rm For the \it next step
\rm we start with the potential $V(x)={6/x^2}$, take the general $E=0$ solution $\phi(x)={\mu/x^2}+x^3$ and construct, as before, the new, interesting potential
\be\la{22}
V(x)={2 \over x^2}{6x^{10}-18\mu x^5+\mu^2 \over (x^5+\mu)^2}
\ee
that we plot in Fig.2

\begin{figure}
\begin{center}
\includegraphics[width=13cm]{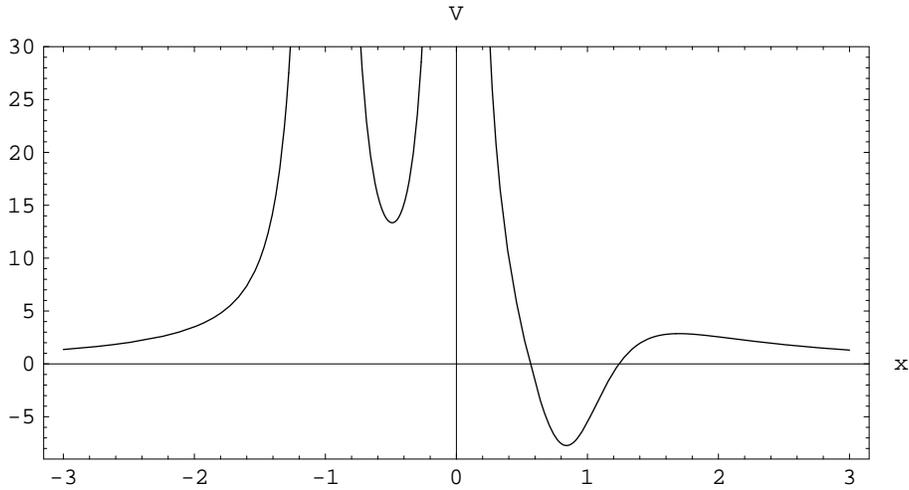}
\end{center}
\caption{{The three regions of the potential (3.22) with ($\mu\equiv1$) }}
\label{Figure 2}
\end{figure}

This potential contains \it three
\rm disconected pieces:

{I}- $x>0$ . Atraction plus repulsion.

{II}- $-c<x<0$ . A confining potential ($c\equiv \mu^{1/5}>0$)

{III}- $x<-c$ . A repulsive potential.

The solutions are again straightforward but tedious, the procedure to obtain them is as in the previous case, and we just indicate and quote the results:

{I}-$x>0$ . There is a \it single bound state
\rm with $k=0$ and (unnormalized) wavefunction
\be\la{23}
\psi_0(x)={x^2 \over \mu + x^5}=\phi^{-1}
\ee
and total reflection with wavefunction
\be\la{24}
\psi_R(x)=A_3A_2A_1\phi_k(x)
\ee
with
\be\la{25}
A_3=D+W'=D+{2\mu-3x^5 \over x(\mu +x^5)}
\ee
and $A_2=D-{2/x}$ , $A_1=D-{1/x}$ . Hence, if $S(k)$ is the S-matrix for the previous $V(x)={6/x^2}$ and $\hat S(k)$ the new one, 
\begin{eqnarray}\la{25.b}
\psi_k(x>>0)&=&(D+W'(\infty))\Phi_k(x>>0)=(D+W'(\infty))(\exp(-ikx)-S(k)\exp(ikx))\nonumber\\
&=& N(\exp(-ikx)-\hat S(k)\exp(ikx)), \ hence \\
\hat S(k)&=& S(k){(W'(\infty)+ik)/(W'(\infty)-ik)}=(+1)(-1)=-1 \nonumber
\end{eqnarray}
because $W'(\infty)=0$ and $S(k)$ , due to ${2(2+1)/x^2}$ , is $=+1$. So
\be\la{26}
\hat S(k)=1 \qquad or \qquad \delta(k)={\pi \over 2} \ mod \ \pi !
\ee

These results are worth commenting: First, the $E=0$ bound state is obvious, because $A\phi=0$ in the previous $V={6/x^2}$ potential implies $A^\dag \phi^{-1}=0$, and $\phi^{-1}$ zero-less, normalizable. Because of the repulsive tail, it is a \it bona fide
\rm bound state, \it not 
\rm an $E=0$ resonance, so it will contribute $+\pi$ to Levinson's theorem \cite{7}.

The constant phase shift is suspitious of \it some kind
\rm of scale invariance. In fact, an scale-invariant bound state can exist if at all, at $E=0$ ; this is our case! The interpretation of the phase shift ``span'' is this: for ${k\to 0}$, the bound state contributes $+\pi$ , the long tail ${12/x^2}$ gives $-{3\pi/2}$ : so $\delta(0)=-{\pi/2}$ , or $S(k=0)=-1$ . At ${k\to \infty}$ , the short tail contributes $-{\pi/2}$ , so $\delta(k=\infty)=-{\pi/2}$ , and the total span of $\delta(k)$ is zero (while it was $+{\pi/2}$ in the previous case).

It is remarkable that a variable (\it i.e.
\rm , not purely centrifugal) potential, indeed supporting a ($E=0!$) bound state is still ``conformal'' and produces constant phase shift. We offer the following explanation:

The previous potential $V_0(x)={6/x^2}$ is manifestly scale invariant:
\be\la{26.b}
[\hat D, H_0]=-2H_0 \quad  where \quad  H_0=-D^2+V_0=A^\dag A\nonumber
\ee
and $\hat D\equiv x\cdot D$ is a dilatation generator. Now
\be\la{26.c}
H_1=AA^\dag =A\cdot(A^\dag \cdot A)\cdot A^{-1}= A\cdot H_0 \cdot A^{-1} \nonumber
\ee
Hence
\be\la{27}
[\hat D_A,H_1]=-2H_1 \ with \ \hat D_A=A\cdot \hat D \cdot A^{-1}
\ee
(Notice $A$ is invertible outside the bound state). Now for $x>>0$ , $W'(\infty)=0$ , so $A={D+W'\to D}$ , and therefore 
\be\la{28}
{\hat D_A\to D(x\cdot D)D^{-1}=x\cdot D +1=\hat D+1}
\ee
That is: the traslated symmetry of the new hamiltonian still guarantees constancy of the phase shift.

To the best of our knowledge, this is a first case of  \it a potential, not purely centrifugal, with constant phase shift.
\rm 

{II}- $-c<x<0$ . This confining potential produces of course an uninteresting, infinite ladder of bound states, reminiscent of the potential $V(x)=-2\cosh^{-2}(x)$ alluded to in Sect. 1. The eigenvalues are $E_n=k_n^2$ where
\be\la{29}
\tan(k_n c)={3k_n c \over 3-(k_n c)^2}
\ee
which is a simple trascendental equation with infinite roots $0<k_1<k_2...<k_n<...$ which tend to $n\pi$ for $n>>1$ ; hence the spectrum is asymptotically parabolic, as for a particle in an infinite box; this is to be expected, as the potential (also in the $V(x)=2\sec^2(x)$ case) is negligible for higher excited wavefunctions. In fact, the normalizable wavefunctions can be written easily, but we refrain of doing it.

{III}- $x<-c$ . At the left, a purely repulsive potential produces only total reflection, and the S-matrix and the phase shift are computed to be
\be\la{30}
S(k)={(kc)^2-3ikc-3 \over -(kc)^2-3ikc+3}, \qquad \tan(\delta + {\pi/2})={-3kc \over (kc)^2-3}
\ee
So the total span $\delta(0)-\delta(\infty)$ is now $=\pi$ , and the phase shift is \it not
\rm constant, going smoothly from $-{3\pi/2}$ to $-{\pi/2}$ in the interval ${k=0\to k=\infty}$. Of course, ``conformal'' invariance has been lost because the singular point is at $x=-c$ , not at $x=0$.

From the many possible generalizations, we consider in this paper just one more case: the general partner of the $n$-step manifest scale invariant potential $V_0(x)={n(n+1)/x^2}$ . The two $E=0$ solutions (both unphysical again) are $x^{n+1}$ and $x^{-n}$ ; so defining
\be\la{31}
\phi(x)={\mu/x^n}+x^{n+1} \qquad \mu>0
\ee
the partner family is
\be\la{32}
V_{\mu}(x)={(n+1)(n+2)x^{4n+2}-6\mu n(n+1)x^{2n+1}+\mu^2 n(n-1) \over x^2(\mu+x^{2n+1})^2}
\ee
which again exhibits the three regions as before. In particular

$x>0$ : a partly attractive potential, which supports again just a \it bound state
\rm at \it zero energy
\rm ; total reflection occurs with (again) \it constant
\rm phase shift. The bound state is $\phi^{-1}$ , of course, and it turns out that 
\be\la{33}
S(k)=(-1)^{n+1}
\ee
by the same argument as before, namely $S(k)=-S_n(k)$ , where $S_n(k)$ is the S-matrix for $V_0(x)={n(n+1)/x^2}$ , namely $S_n(k)=(-1)^n$.

We have therefore found and infinite family of ``scale'' invariant potentials, with a unique $E=0$ normalizable bound state, and constant (in fact $\pm 1$) S-matrix. The (modified) Levinson theorem applies; namely the span $\delta(0)-\delta(\infty)$ is zero: at low $k$, there is a $+\pi$ contribution from the bound state, and $-{n(n+1)\pi/2}$ value from the long tail. At large $k$ , the short tail takes over, contributing ${-(n-1)\pi/2}$ . The constancy of $\delta(k)$ comes, as before, from the appropiate conjugation of the manifest dilatation symmetry of the previous potential, just as in the worked-out case $n=2$.

In the confining region $-c<x<0$ , with $c=+\mu^{1/(2n+1)}>0$ , there is a pure point spectrum, with again a limiting parabolic growth in the energy. The spectral equation is a natural generalization of (\ref{29}); we state only the next case, $n=3$; the trascendental eigenvalue equation is
\be\la{34}
\tan(kc)={7(kc)^3-105(kc) \over 42(kc)^2+105}
\ee

Finally, in the pure repulsive part of the potential, $x<-c$ , there is only total reflection with a simply variable phase shift. The total span is 
\be\la{35}
\delta(0)-\delta(\infty)=-{(n+1)\pi/2}-(-{\pi/2})=-{n\pi/2}  
\ee
because the potential behaves like $+{2/(x+c)^2}$ close to the pole.
The exact S-matrix can be calculated as before. We just quote the result only again for $n=3$ :
\be\la{36}
S(k)={42(kc)^2+105-i(7(kc)^3-105(kc)) \over (complex \  conjugate)}
\ee

The general $S(k)$ starts at $S(0)=+1$ for $n=3,5,7,...$ , and $S(0)=-1$ for $n$ even; it becomes $S(\infty)=-1$ after $n$ half-turns. The phase shift connects smoothly $-{(n+1)\pi/2}$ at $k=0$ with $-{\pi/2}$ at $k=\infty$ .

We can see also why the first case $n=1$ is special: at right the potential is $+{6/x^2}$ for $n=2$ , and at $x=0$ is $V=0$ , as $n-2=0$ so in this case there are only two regions with \it no
\rm confining part. 
\section{Other potentials}
\rm Once the general procedure is understood, is a matter of mechanical calculations to find and to solve any other $V=0$-related potentials. We shall report on a full investigation elsewhere \cite{11}

Here we just report that we can, by our procedure, recover many of the ``shape invariant'' potentials in the review Infeld-Hull paper \cite{1}
; in fact, all the families included in the ``A-type'' classification of \cite{1}
 . The other types B...I are in some way degenerate: they include, among others, the oscillators, Kepler and Morse potentials, which are \it not directly
\rm connected to the $V=0$ case, but still are ``shape invariant'' and solvable. As shown in \cite{11}
, the Kepler problem is related to the $V=0$ potential in a constant curvature (spherical for bound states) space.

The natural minimal generalization  of $V={2/x^2}$ is obviously the centrifugal potential 
\be\la{37}
V(x)={n(n+1) \over x^2} \qquad n=0,1,2...
\ee
This is obtained from $V=0$ by making use of the solutions $x, x^2, x^3,...,x^n$ in each step.

The minimal natural extension of $V=2\sec^2(x)$ is
\be\la{38}
V(x)=+n(n+1)\sec^2(x) \qquad n=0,1,2...
\ee

The intertwining superpotential satisfies $W'_n(x)=n\tan(x)$ , with $\phi_n(x)=\cos^n(x)$ as the generating wavefunction.; notice the energy scale gets displaced; this confining-potential family contains a pure discrete spectrum, approaching the parabolic infinite-box situation.

Similarly
\be\la{39}
V(x)=-n(n+1)\cosh^{-2}(x) \qquad n=0,1,2...
\ee
comes from $W'_n(x)=-n\tanh(x)$ and  $\phi_n(x)=\cosh^n(x)$.

There are $n$ bound states and a $E=0$ resonance, plus perfect transmission (no reflection); it is the well known ``$n$-solitonic'' potential, with all the elementary solitons on top of each other at $x=0$ \cite{5}
.

The final minimal family is
\be\la{40}
V(x)=+n(n+1)\sinh^{-2}(x) \qquad n=0,1,2,...
\ee
This comes from $W'_n(x)=+n\coth(x)$ and $\phi_n(x)=\sinh^n(x)$ . It corresponds to total reflection, with variable phase shifts, and no bound states.

All these four families are still exactly solved even for  ${n\to \lambda}$ noninteger, by ``prolongation'' (see \cite{1} or the review \cite{8})   
; they are shape-invariant \cite{12}
 and therefore included in \cite{1}
. As they are \it not
\rm related \it directly
\rm with the vacuum $V=0$ case, we do not discuss them. 

The only potential of ``A'' type of \cite{1}
 not include so far is
\be\la{41}
V(x)={a+b\cos(x) \over \sin^2(x)}
\ee
This can still be also obtained in our scheme in the following, indirect way: the potential
\be\la{42}
V(x)={{3/4} \over \sin^2(x)}
\ee
is a prolongation of the $V_1(x)=2\csc^2(x)$ of $\S$ 1, and it admits the \it unphysical
\rm eigenfunction
\be\la{43}
\phi(x)=\sqrt{\sin(x)} \cot({x/2})
\ee
Hence the corresponding partner potential is, with $W'(x)=-\cot({x/2})+\csc(x)$ ,
\be\la{44}
V_1(x)={{7/4}-2\cos(x) \over \sin^2(x)}
\ee
which is of type (\ref{41}). The energy of the unphysical solution $\phi$ is $+{1/4}$ , whereas the ground state of (\ref{44}) is $\sin^{3/2}(x)$ , with energy $=+{9/4}$ . We conclude that the ``A'' type family of Infeld-Hull \cite{1}
 can be included also in our scheme of things.
\section{Conclusion}

\rm The whole set of analytically soluble potentials (not to speak of the quasi-soluble ones \cite{13}
) is very, very large. In this paper we have shown how starting with the free case, $V(x)=0$ , and just by playing around with the unphysical solutions for $E=0$ only, a large family is obtained;  the generic case includes a confining potential defined in a segment of the line, a purely repulsive half-line defined potential, and an also half-line defined potential, supporting a \it bona fide
\rm unique $E=0$ bound state with trivial (i.e. constant $=\pm 1$) S-matrix.

The natural generalization of the four different potentials obtained in the first step from $V(x)=0$ includes all the non-degenerate cases in the Infeld-Hull series, if we include prolongations, that is, substituting $n(n+1) \ n\in N$, by $\lambda(\lambda+1)$ for arbitrary, real positive $\lambda$. They correspond to solutions of the hypergeometric equation, which is also related to the $SL(2,R)$ group.
; the degenerate I-H cases B...I (i.e. Coulomb,...) correspond to solutions of the \it confluent
\rm hypergeometric equation.

There is still work in progress; we have \it not
\rm exhausted even the $E=0$ family (for example, we can iterate the potential (\ref{32})!). As stated, we plan to report on other cases in a later publication; see also \cite{11}

The two Appendices explain the Darboux method and elaborate on the KdV connection, as promised.
\section*{Acknowledgements} \nonumber

\rm One of us (LJB) started this work some time ago in collaboration with A.~ del Sol, and we want to thank him for some early calculations. Discussions with Prof.~ M.~ Asorey and Dr.~ A.~ Ramos in our Department were clarifying. We have been partially supported by a grant MCyT FPA-2.000 \# 1252.
\setcounter{section}{0}
\renewcommand{\thesection}{\Alph{section}}
\renewcommand{\theequation}{\thesection.\arabic{equation}}
\section{Appendix} \setcounter{equation}{0}  

We probe the results used in \S 1. From $H\Psi = (-D^2+V)\Psi =E\Psi$, define 
\be
\exp(-W(x))\equiv \Psi
\ee
$W$ satisfies a Riccati first order equation $V-E=W'^2-W''$: the scale invariance ${\psi \to \lambda \psi}$ becomes translation invariance for $W$, hence $W$ itself does not appear in the new equation. $\psi$ \it needs not
\rm to be physical, \it i.e. 
\rm it can be singular. But now the hamiltonian factorizes:
\be
-D 2+W'^2(x)-W''(x)=(-D+W'(x))(D+W'(x))=A^\dag A
\ee
with $A=D+W'(x)$. We obtain
\be
A^\dag A\Psi=(H-E)\Psi
\ee

The partner hamiltonian is defined as $H'=AA^\dag +E$, so 
\be
H\phi'=E'\phi'=(A^\dag A+E)\phi' \quad \Rightarrow \quad (AA^\dag +E')A\phi'=E'A\phi'
\ee

For each solution $(\phi',E')$ of the former $H$ we obtain a solution $(A\phi',E')$ of the new $H'$. Of course, $\phi'$ might be physically unacceptable. This is the essence of the method.

Now for the second solution. $\phi=\exp(-W)$ implies $A\phi=0$, $A=D+W'$. Or $(\exp(-W)\cdot D \cdot \exp(+W))\exp(-w)=0$, hence $(\exp(+W)\cdot D \cdot \exp(-W))\phi^{-1}=0$, $A^\dag \phi^{-1}=0$ so $\phi^{-1}$ corresponds to $\phi$ for $E'=E$. Now if $\phi'$ is the second solution of the original $H$ with energy $E$, $A^\dag A\phi'=0$ but $A\phi'\not =0$, hence $A\phi'$ is in the kernel of $A^\dag$, and therefore
\be
\exp(-W)\cdot D \cdot \exp(+W) \phi'=\exp(W), \quad or \quad \phi'=\phi\int \phi^{-2}\,dx
\ee
as stated.

The use of the second solution to generate new potentials seems to start with \cite{9}
; see also the previous work of Abraham and Moses \cite{10}  
\section{Appendix} \setcounter{equation}{0} 

The KdV equation ($u=u(x,t), \ U_{,t}={\partial u \over \partial t}$ etc.)
\be
u_{,t}=6uu_{,x}-u_{,xxx}
\ee
is one of the deformation equations associated to the Schr\"odinger equation, $H(\mu)= -D^2+u(x,\mu)$, where $\mu (=t)$ is the deformation parameter. For this reason some simple solutions of KdV are interesting potentials for the linear problem; we take some results from \cite{5}
.  The travelling wave solution $u=u(x-vt)$
\be
u(x,t)=-({1/2})v\cosh^{-2}(\sqrt{v} (x-vt-x_0))
\ee
corresponds to our first step with $E<0$ and $\phi=\cosh(x)$; it is the solitonic potential. Multisolitonic potentials correspond to iteration from this solution, but these are not considered in this paper.

Rational solutions of KdV are closer to our potentials; for example $u={2/x^2}$ arises as the simplest t-independent rational solution, and it is our first potential from the $E=0$, $\phi=x$ solution. The natural scaling invariance of KdV  ${x\to \lambda x}, \ {t\to \lambda^3 t}, \ {u\to \lambda^{-2} u}$ leads at once to the rational solution
\be 
u(x,t)={6x(x^3-24t) \over (x^3+12t)^2}
\ee
which is our potential (\ref{10}) (with $\mu=12t$). Similarly the other rational potentials we obtain are connected with rational solutions of higher-order KdV-hierarchy equations; we shall report on a full investigation elsewhere.

\end{document}